\documentclass[hyper,12pt,drafts]{JHEP3}

\usepackage[centertags]{amsmath}
\usepackage{amsfonts,amsmath,amssymb,epsf}
\usepackage{graphicx}
\usepackage[final]{showkeys}

\newcommand{\half}{\frac{1}{2}}

\newcommand\nn{{\nonumber}}
\newcommand{\beq}{\begin{equation}}
\newcommand{\eq}{\end{equation}}

\def\bea{\begin{eqnarray}}
\def\ea{\end{eqnarray}}

\def\a{\alpha}

\title{On the IR completion of geometries with hyperscaling violation}

\author{Jyotirmoy Bhattacharya$ ^{\, \diamondsuit}$\footnote{\tt{jyotirmoy.bhattacharya@ipmu.jp}}, \,
Sera Cremonini$ ^{\,\clubsuit,\spadesuit}$\footnote{\tt{s.cremonini@damtp.cam.ac.uk}} \,
and Annamaria Sinkovics$ ^{\,\clubsuit}$\footnote{\tt{A.Sinkovics@damtp.cam.ac.uk}}\\
\it $ ^\diamondsuit$ \textit{Kavli Institute for the Physics and Mathematics of the Universe (WPI),
The University of Tokyo, Kashiwa, Chiba 277-8583, Japan}\\
\it $ ^\clubsuit$
\it DAMTP, Centre for Mathematical Sciences, University of Cambridge, Wilberforce Road, Cambridge, CB3 0WA, UK \\
\it $ ^\spadesuit$ George and Cynthia Mitchell Institute for Fundamental Physics and Astronomy,
\it Texas A\&M University, College Station, TX 77843--4242, USA}
\abstract{
We study solutions to Einstein-Maxwell-dilaton gravity with a constant magnetic flux which describe, in the holographic AdS/CFT framework,
field theories characterized by a dynamical critical exponent and a hyperscaling violation exponent.
Such solutions are known to be IR-incomplete due to the presence of a running dilaton, which drives the theory towards strong coupling in the IR,
where quantum corrections become important.
After introducing generic corrections, in this note we examine the conditions for the emergence of an
$AdS_2 \times R^2$ region close to the horizon, which provides an IR-completion for the hyperscaling violating solutions.
In the presence of these corrections, we construct explicit numerical solutions where the geometry flows from $AdS_4$ in the UV to $AdS_2 \times R^2$ in the deep IR, with an
intermediate region which exhbits both hyperscaling violation and Lifshitz-like scaling.
 We also provide constraints on the structure of Einstein-Maxwell-dilaton theories that admit
 such solutions, as well as an emergent $AdS_2 \times R^2$ region in the infrared.
}
\preprint{MIFPA-12-30, \\ DAMTP-2012-56, \\IPMU12-0150.}
\begin{document}
\maketitle

\section{Introduction} \label{intro}

In recent years much of the focus of the holographic gauge/gravity duality has shifted towards geometries which exhibit interesting scaling
properties, which have shown to be a rich playground for constructing toy models of condensed matter systems.
A prime example is that of Lifshitz metrics, dual to field theories that violate Lorentz invariance,
\beq
\label{LifMetric}
ds^2_{d+2} = - \frac{1}{r^{2z}} dt^2 + \frac{1}{r^2} \left( dr^2 +  d\vec{x}^2 \right)\, ,
\eq
which are parametrized by a dynamical critical exponent $z$ and are invariant under
\beq
t \rightarrow \lambda^z t \, , \quad r \rightarrow \lambda r \, , \quad  x_i \rightarrow \lambda x_i \; . \nn
\eq
Metrics of the form (\ref{LifMetric}) are exact solutions to gravitational theories coupled to a matter sector \cite{Kachru:2008yh,Taylor:2008tg},
with an abelian gauge field and a dilaton providing the simplest realization of the latter.

It was realized more recently (see e.g. the work of
\cite{Gubser:2009qt,Cadoni:2009xm,Charmousis:2010zz,Perlmutter:2010qu,Iizuka:2011hg,Gouteraux:2011ce,Huijse:2011ef})
that Einstein-Maxwell-dilaton theories
\beq
\label{IntroLag}
{\mathcal L}_{d+2} = R-2 \, (\partial \phi)^2 - f(\phi) F_{\mu \nu} F^{\mu \nu} - V(\phi)
\eq
can support quite generally -- for simple choices of gauge kinetic function and scalar potential --
larger classes of scaling solutions
exhibiting an additional exponent $\theta$,
\beq
\label{HVMetricIntro}
ds^2_{d+2} = r^{-\frac{2(d-\theta)}{d}} \left( - r^{-2(z -1)} dt^2 + dr^2  +  d\vec{x}^2 \right) \, ,
\eq
supported by a running scalar field
\beq
\label{phiintro}
\phi(r) \sim \log(r) \, .
\eq
In particular, it was emphasized in \cite{Huijse:2011ef} that theories with $\theta \neq 0$ realize
systems with \emph{hyperscaling violation} in the dual field theory.

While they are not scale invariant, the metrics (\ref{HVMetricIntro}) are conformal to Lifshitz spacetimes (\ref{LifMetric}),
and exhibit the scaling
\beq
t \rightarrow \lambda^z t \, , \quad r \rightarrow \lambda r \, , \quad  x_i \rightarrow \lambda x_i \, ,
\quad ds \rightarrow \lambda^{\theta/d} ds \; . \nn
\eq
While in systems which preserve hyperscaling the free energy scales by its naive dimension, so that
$s\sim T^{d/z}$, a non-zero $\theta$
modifies the scaling of the entropy density\footnote{Note that there are cases in which, after uplifting to higher dimensions,
one regains the expected  `naive' scaling of thermodynamic quantities, thus explaining
 the unusual lower-dimensional behavior \cite{Gouteraux:2011ce}. In such cases the higher-dimensional embedding also offers
 a possible resolution of the singular horizon behavior of the lower-dimensional (zero temperature) solutions.},
leading to $$s \sim T^{\frac{d-\theta}{z}} \, .$$
Precisely for this reason, scaling geometries with $\theta =d-1$ have been relevant
for probing compressible states with hidden Fermi surfaces, for which $s\sim T^{1/z}$ in general dimensions.
Hyperscaling violating solutions have also been of interest for their
connection with log violations of the area law of entanglement entropy \cite{Ogawa:2011bz}.
We refer the reader to e.g. \cite{Kulaxizi:2012gy,Ammon:2012je,Cadoni:2012uf,Perlmutter:2012he,Dey:2012rs,Dey:2012tg,Narayan:2012hk,Hartnoll:2012wm,Dong:2012se,Li:2010dr,Li:2009pf}
for various properties of these systems, and for ways to obtain them within supergravity and string theory embeddings.

The running of the scalar (\ref{phiintro}) in this class of geometries is a reflection of the fact that the solutions
cannot be trusted in the deep IR -- they are `IR incomplete' -- and should only be thought of as being an accurate description of the geometry
in some intermediate near-horizon region.
For magnetically charged branes -- the case we are interested in here -- the break-down of the solutions
results from the fact that the dilatonic scalar runs towards \emph{strong coupling} near the horizon --
the low-energy theory itself is breaking down, and quantum effects (in a putative string realization) are no longer negligible.
For their electrically charged cousins, on the other hand, the dilaton drives the system to extreme weak coupling close to the horizon, and
$\alpha^\prime$ corrections are expected to become important. While the fate of these geometries is equally interesting,
in this note our focus will be on understanding the behavior of the strongly coupled, magnetic case (see \cite{Horowitz:2011gh} and
\cite{Bao:2012yt} for related discussions in the context of Lifshitz systems without running couplings).

For the case of a brane exhibiting Lifshitz scaling, which is generated by
\beq
f(\phi) \propto e^{\, 2\alpha\phi}  \quad \quad \text{and} \quad \quad V(\phi) = - 2\Lambda \, ,
\eq
with $\alpha$ dictating the strength of the Lifshitz dynamical exponent $z$,
this point was noted in \cite{Goldstein:2009cv,Goldstein:2010aw} and was the focus of the analysis of \cite{Harrison:2012vy},
where the inclusion of (a toy model for) quantum corrections was shown to lead to a modification of the geometry in the deep IR
and the appearance of an $AdS_2 \times R^2$ description, thus IR completing the running dilaton solutions.

In this note, we would like to extend the analysis of \cite{Harrison:2012vy} to the more general class of
geometries exhibiting both hyperscaling violation and Lifshitz-like scaling, which are generated by
\beq
\label{SpecificfV}
f(\phi) \propto e^{\, 2\alpha\phi}  \quad \quad \text{and} \quad \quad V(\phi) \propto e^{- \eta \phi} \, ,
\eq
with the parameters $\alpha$ and $\eta$ determining the scaling exponents $\theta$ and $z$.
Magnetically charged brane solutions in this theory also contain a running dilaton --
leading to the same issue of strong coupling at the horizon --
and therefore suffer from the same `IR-incompleteness' discussed above.

With these motivations in mind, our goal here is to probe the IR fate of this class of geometries, working in
particular with solutions that are asymptotically AdS.
To this end, we will follow the strategy of \cite{Harrison:2012vy} and consider a toy model for the quantum-corrected version
of the theory, by appropriately modifying the structure of the gauge kinetic function.
Specifically, we will mimic the effects of quantum corrections by elevating
$f(\phi)$ to an expansion in powers of the coupling $g \equiv e^{-\alpha\phi}$,
\beq
\label{ToyCorrections}
f(\phi) = e^{2\alpha\phi} \quad  \longrightarrow \quad e^{2\alpha\phi} + \xi_1 + \xi_2 e^{-2\alpha\phi} + \ldots
\eq
and more generally by replacing it with an arbitrary function, $f(\phi) \rightarrow e^{2\alpha\phi} + \mathcal{G}(\phi)$.
Corrections such as (\ref{ToyCorrections}),
which become stronger and stronger as the deep IR is approached, will generate
-- in appropriate regions of parameter space -- an effective potential for the scalar, stabilizing it at a constant value at the horizon.
Thus, as in \cite{Harrison:2012vy}
we will see the emergence of an $AdS_2 \times R^2$ region very close to the horizon, providing an IR-completion to the scaling geometry
which would not have been generically possible at the classical level\footnote{For the class of theories we are studying here,
described by (\ref{IntroLag}) and (\ref{SpecificfV}) in the presence of a constant magnetic field,
we will find that $AdS_2 \times R^2$ is only possible classically for the special case $\theta =2$ (and z finite),
or alternatively in the $z\rightarrow\infty$ limit.}
for (\ref{SpecificfV}).
These geometries will flow, then,
from $AdS_2 \times R^2$ in the deep IR
to $AdS_4$ in the UV, traversing during the flow an intermediate region which exhibits both hyperscaling violation and Lifshitz-like scaling.

Finally, we note that the requirement of the existence of an $AdS_2$ factor in the infrared
places simple restrictions on the structure of the arbitrary correction $\mathcal{G(\phi)}$
to the gauge kinetic function --
constraining in particular its value and slope at the horizon.
These restrictions are valid independently of the origin of the corrections -- and in particular,  of whether
they are classical\footnote{It is plausible that a gauge kinetic function of the form of (\ref{ToyCorrections})
may be realized even classically within a consistent supergravity truncation, with $\xi_1$ and $\xi_2$ set to fixed values.
Our construction would still be applicable in such cases.} or quantum in nature.
Similar considerations can also be easily applied to generic corrections to the scalar potential.
Thus, constraints of this type may offer insight into the broader question of the emergence of hyperscaling violation
in the \emph{intermediate} region of a  solution to classical Einstein-Maxwell-dilaton theories
with more generic choices of $V(\phi)$ than those studied here.
It may also prove useful for realizing flows of the type described above in the context of supergravity and string theory constructions.
Before we conclude, we would like to point out that by turning on a (small) magnetic field in addition to an electric field \cite{Kundu:2012jn},
the action with (\ref{SpecificfV}) can support a near-horizon $AdS_2 \times R^2$ already classically --
for appropriate choices of parameters, the electric and magnetic charge contributions
to the effective potential lead to a stabilization of the scalar, providing an
explicit \emph{classical} realization of the mechanisms we have just discussed.
Thus, the analysis in \cite{Kundu:2012jn} leads to a very interesting picture which is complementary to that studied in this note, where
we have not allowed for any electric flux.

The structure of this note is as follows.
In \S\ref{setup} we introduce our setup, that of Einstein-Maxwell-dilaton theory, and discuss the class of solutions we will focus on.
We also derive a set of simple constraints that a generic gauge kinetic function and scalar potential must satisfy,
in order to obtain metrics with hyperscaling violation and Lifshitz-like scaling.
In \S\ref{s:ads2} we discuss under which conditions the theory admits an $AdS_2 \times R^2$ description in the deep IR,
in the presence of a class of quantum corrections to the action.
In this section we also setup the irrelevant perturbations which will take the geometry away from the IR, and describe
conditions for the existence of $AdS_4$ in the UV.
Finally, in \S\ref{num} we analyze numerically the entire flow from the near-horizon $AdS_2 \times R^2$ region to the boundary.
We start by discussing the case in which the UV geometry is hyperscaling violating, and then move on to the case in which the
geometry approaches $AdS_4$.
In Appendix \ref{app:nullcond} we present a short study of the null energy condition
for the classical theory which gives rise to the hyperscaling violating solutions.

{\it Note added in v1}: While we were completing this work we became aware  of \cite{Kundu:2012jn} where similar results have been obtained.

\section{The Setup -- Einsten-Maxwell-dilaton theory}\label{setup}

Our starting point is Einstein-Maxwell-dilaton theory,
\beq
\label{GeneralLag}
{\mathcal L}_{d+2} = R-2 \, (\partial \phi)^2  - f(\phi) F_{\mu \nu} F^{\mu \nu} - V(\phi) \, ,
\eq
where we denote by $D=d+2$ the total dimensionality of the space-time.
As we already discussed in the introduction, theories of this type have a rich structure, and give rise to
geometries which exhibit interesting scaling properties.
In particular, by appropriately choosing the gauge kinetic function $f(\phi)$ and the scalar potential $V(\phi)$,
one can engineer metrics of the form\footnote{Here we are following the notation of \cite{Dong:2012se}.}
\beq
\label{HVMetric}
ds^2_{d+2} = r^{-\frac{2(d-\theta)}{d}} \left( - r^{-2(z -1)} dt^2 + dr^2  +  d\vec{x}^2 \right) \, ,
\eq
characterized by two independent exponents, the Lifshitz critical exponent $z$
and the hyperscaling violation exponent $\theta$ \cite{Charmousis:2010zz, Huijse:2011ef}.

In this note, we are interested in exploring (extremal) solutions to (\ref{GeneralLag}) which are \emph{magnetically charged},
with the goal of gaining insight into how they behave in the deep IR, as the theory runs towards strong coupling.
In this section, we will start by deriving the types of constraints on the structure of \emph{generic} functions $f(\phi)$ and $V(\phi)$
needed to obtain metrics of the form of (\ref{HVMetric}),
which exhibit both Lifshitz-like scaling and hyperscaling violation.
This will serve as motivation for using a gauge kinetic function and scalar potential of the form of (\ref{SpecificfV}),
and will give us an explicit map between the lagrangian parameters $\{\alpha,\eta\}$ and the exponents $\{z,\theta\}$.
We will then briefly summarize the main properties of the solutions to this system, which are used throughout the analysis.

From now on we will restrict our attention to four dimensions, taking $d=2$.
We choose the background gauge field to be that of a constant magnetic field,
\beq
F = Q_m \, dx \wedge dy \; ,
\eq
and parametrize the metric, which we take to be homogeneous and isotropic, by
\beq
\label{abcAnsatz}
ds^2 = L^2 \left( -a(r)^2 dt^2 + \frac{dr^2}{a(r)^2} + b(r)^2 d\vec{x}^2 \right) \, .
\eq
Einstein's equations for this theory are given by
\begin{equation}
 R_{\mu \nu} + \half \left(V(\phi) - R\right) g_{\mu \nu} = 2 ~\partial_{\mu} \phi \partial_{\nu} \phi - g_{\mu \nu} ~\partial_{\rho} \phi \partial^{\rho} \phi
- 2 f(\phi) \left( F_{\mu \rho} F^{\rho}_{\ \nu}- \frac{1}{4} g_{\mu \nu} F_{\rho \sigma} F^{\sigma \rho}\right)  ,
\end{equation}
while the scalar equation and Maxwell's equations take the simple form
\bea
&& 4 D_{\mu} \partial^{\mu} \phi - f_\phi(\phi) F_{\mu \nu} F^{\mu \nu} - V_\phi(\phi) = 0 \, , \\
&&  D_{\mu} \left( f(\phi) F^{\mu \nu} \right) = 0 \, ,
\ea
where $D$ denotes the covariant derivative with respect to the Levi-Civita connection, and we have defined
$f_\phi \equiv \partial_\phi f$ and $V_\phi \equiv \partial_\phi V$.
After simple manipulations, this system of equations
can be easily shown to reduce to
\bea
\label{aEOM}
\phi^{\, \prime \, 2} &=&  -\frac{\; b^{\, \prime\prime}}{b} \, , \\
\label{cEOM}
4b^2 L^2 V(\phi) &=& -2 \, (a^2 b^2)^{\, \prime\prime}  \, , \\
\label{bEOM}
f(\phi) Q_m^2 - 2 b^4 L^4 V(\phi) &=& 2 \, L^2 b^2 \left(b^2 (a^{2})^\prime \right){\, ^\prime} \, , \\
\label{phiEOM}
f_\phi(\phi) \, Q_m^2 + 2 \, b^4 L^4 \, V_\phi(\phi) &=& 8 \, L^2 b^2  \left(a^2 b^2 \phi^{\, \prime}\right)^{\, \prime} \, ,
\ea
where primes denote radial derivatives, i.e. $^\prime \equiv \partial_r$, and we have already used our flux ansatz.

\subsection{Engineering Hyperscaling Violation}\label{enghyper}

Here we would like to address the question of what type of constraints the requirements of
hyperscaling violation and Lifshitz-like scaling place on the structure of generic $f(\phi)$ and $V(\phi)$.
With this in mind, we start by taking the metric (\ref{abcAnsatz}) to be
\beq
\label{HyperAnsatz}
a(r) = C_a \, r^{1-\gamma} \, , \quad \quad  b(r) = C_b \, r^{\, \beta} \, ,
\eq
characterized by two `scaling exponents' $\beta$ and $\gamma$. Note that in our metric \eqref{abcAnsatz}
we have a slightly different choice of gauge compared to \eqref{HVMetric}. By a suitable redefinition
of the radial coordinate
\beq
\label{mapr}
r \quad \rightarrow \quad r^{\frac{1}{\gamma- \beta}}\, ,
\eq
our metric can be mapped to \eqref{HVMetric},
where $z$ and $\theta$ in \eqref{HVMetric} are related to $\beta$ and $\gamma$ through the following relations,
\beq
\label{betagamma}
\beta = \frac{\theta-2}{2(\theta-z)} \quad \quad \text{and} \quad \quad \gamma = \frac{\theta}{2(\theta-z)} \, .
\eq
Thus, the parameter $\gamma$ directly measures the strength of the violation of hyperscaling.
As we will discuss in more detail below, note that when $\theta=\gamma=0$
the system reduces to the Lifshitz-like scaling case, as it should. In particular, we recover the relation
$\beta=1/z$ familiar from studies of Lifshitz solutions.

Plugging the power-law ansatz (\ref{HyperAnsatz}) into the equations of motion, we note first that (\ref{aEOM}) reduces to the simple form
\bea
\label{aEOMHyper}
\phi^{\, \prime \, 2} &=&  \frac{\beta-\beta^2}{r^2} \, ,
\ea
from which we can immediately read off that the scalar must run logarithmically\footnote{We are setting the integration constant to zero.},
\beq
\phi(r) =  K \log(r) \, , \quad K^2 = \beta-\beta^2  \, ,
\eq
as expected.
The scalar potential can be extracted from (\ref{cEOM}), and is given by
\beq
\label{HyperScalar}
V = - V_0 \, e^{-\eta\phi} \, ,
\eq
where we have introduced
\beq
V_0 \equiv \frac{C_a^2 (1+ 2\beta - 2\gamma)(1+\beta-\gamma)}{L^2}  \quad \quad
\text{and} \quad \quad  \eta \equiv \frac{2\gamma}{K}  \, .
\eq
Plugging the latter into (\ref{bEOM}), we fix the form of the gauge kinetic function, which is given by
\beq
f(\phi) = e^{\, 2\alpha \phi} \, \frac{2 \, C_a^2 \, C_b^4 \, L^2 (1-\beta-\gamma)(1+2\beta-2\gamma) }{Q_m^2}  \, ,
\eq
where we have introduced an additional parameter
\beq
\alpha \equiv \frac{2 \beta-\gamma}{K} \, .
\eq
For our metric and scalar ansatz, the remaining equation (\ref{phiEOM}) is automatically satisfied by the potential and gauge
kinetic function found above.

In summary, what we have just seen is that the
requirement of a metric which exhibits both anisotropic Lifshitz-like scaling and hyperscaling violation
forces the scalar potential and the gauge kinetic function to be single exponentials, and the dilatonic scalar to run logarithmically,
\beq
f (\phi) = c_1  e^{\, 2\alpha\phi}  \, , \quad  \quad V (\phi) = - V_0  e^{- \eta \phi} \, , \quad  \quad \phi (r) = K \log(r) \, ,
\eq
with the various parameters in the lagrangian $\{c_1, V_0, \alpha, \eta\}$ as well as $K$
directly sensitive to the scaling exponents $\beta$ and $\gamma$ and the constants $C_a,C_b$ and $L$,
\bea\label{earlyhyper}
\alpha  &=& \frac{2 \beta-\gamma}{K} \, , \quad \quad \quad \quad
\eta =  \frac{2\gamma}{K} \, , \quad \quad \quad \quad   K^2 = \beta-\beta^2 \, ,  \nn \\
c_1 &=& \frac{2 \, C_a^2 \, C_b^4 \, L^2 (1-\beta-\gamma)(1+2\beta-2\gamma) }{Q_m^2}  \, , \nn \\
V_0 &=& \frac{C_a^2 (1+ 2\beta - 2\gamma)(1+\beta-\gamma)}{L^2}  \, .
\ea
As a consistency check, we note that this analysis agrees with that of \cite{Iizuka:2011hg}, who considered the (entirely analogous) case of
electrically charged solutions (see also \cite{Cadoni:2011nq}).
In addition, here we have recast the analysis explicitly in terms of the scaling exponents $z$ and $\theta$,
thanks to the relations ({\ref{mapr}) and (\ref{betagamma}).

Clearly, once the theory is specified so that the precise form of $f(\phi)$ and $V(\phi)$ is known,
the structure of the solution is fully determined.
In particular, normalizing the gauge kinetic term so that $f(\phi)=e^{\, 2 \alpha\phi}$ and choosing $V_0=\frac{1}{L^2}$
fixes the ratio\footnote{Although the constant $C_b$ is redundant and could be set to one, leaving it arbitrary turns out
to be useful in our numerical analysis.}  $Q_m/C_b^2$ and the value of $C_a$,
\beq
\label{Qm2C}
Q_m^2 = 2L^2 \, C_b^4 \frac{1-\beta-\gamma}{1+\beta-\gamma}   \quad \quad
\text{and} \quad \quad
C_a^2 = \frac{1}{(1+\beta-\gamma)(1+2\beta-2\gamma)} \, .
\eq
Note that the requirement that $K$ is real forces $0<\beta<1$, which happens to be one of the constraints that follow
from the null energy condition (see Appendix \ref{app:nullcond}).
Also, since in this note we are interested in solutions which run towards strong coupling\footnote{Recall that the
coupling $g = e^{-2\a \phi} = e^{-2(2\beta-\gamma)\log(r)}$ needs to grow as $r\rightarrow 0$.} in the IR, we will always require
$\alpha K >0$, for which we need to have $\gamma < 2\beta$.
The range of $\gamma$ can be refined further by ensuring that the right-hand sides of (\ref{Qm2C})
are positive, i.e. the reality of $Q_m$ and $C_a$. We will come back to these points in the Appendix, when we discuss the null-energy condition.

\subsubsection{Lifshitz as a special case}

We can easily recover the Lifshitz-like scaling, with no hyperscaling violation, by taking $\theta=0$,
which amounts to setting $\gamma=0$ in the conditions above.
While the scalar remains of the logarithmic form, the potential (\ref{HyperScalar}) now becomes a (negative cosmological) constant,
\beq
V = - \frac{C_a^2 (1+2\beta)(1+\beta)}{L^2} \, .
\eq
The gauge kinetic function also remains of the exponential form,
\beq
f(\phi) = e^{\,2\alpha \phi} \, \frac{2(1+\beta-2\beta^2)C_a^2 C_b^4 L^2 }{Q_m^2} \, ,
\eq
where we now have $\alpha = \frac{2 \beta}{K}$ with $K^2 = \beta-\beta^2$ as before, so that
\beq
z = 1 + \frac{4}{\alpha^2} \, .
\eq
Taking the scalar potential to be $V= - 1/L^2$, and normalizing the gauge kinetic function so that it takes the simpler form $f = e^{2\alpha\phi}$,
we find
\beq
C_a^2 = \frac{1}{(2\beta+1)(\beta+1)} \, , \quad \quad Q_m^2 = 2L^2 \,
\frac{1-\beta}{1+\beta}  \, ,
\eq
where for simplicity we have set the redundant constant $C_b$ equal to one.
As a simple check of our results, we note that the form of this solution is in agreement with the analogous one in \cite{Goldstein:2010aw}.

\section{Construction of IR and UV geometry}
\label{s:ads2}

In the previous section we have seen that a system with non vanishing $\theta$ and $z$ can be \emph{engineered
holographically} by working within the framework of Einstein-Maxwell-dilaton theories,
provided the gauge kinetic function and scalar potential are of the form
\beq
\label{fVform}
f(\phi) = e^{2\alpha\phi} \, , \quad V(\phi) = - V_0 e^{-\eta\phi} \, ,
\eq
with the lagrangian parameters $\alpha$ and $\eta$ dictating the form of $z$ and $\theta$.
In particular, magnetically charged solutions to this system exhibiting both Lifshitz-like scaling and hyperscaling violation are of the form
\bea
\label{SolRepeat}
a(r) &=& C_a \, r^{1-\gamma} \, , \quad  b(r) = C_b \, r^\beta \, , \quad \phi(r) = K \log(r) \, ,\\
\beta &=& \frac{ (2\alpha+\eta)^2 }{16+(2\alpha+\eta)^2}\, , \quad \gamma = \frac{ 2\eta (2\alpha+\eta) }{16+(2\alpha+\eta)^2} \, ,\label{SolRepeat2}
\ea
with the remaining constants given in terms of $\{\alpha,\eta\}$ by
 \begin{equation}\label{SolRepeat3}
 \begin{split}
  C_a &= \frac{1}{2} \left( \frac{L^2 V_0 \left(4 \alpha ^2+4 \alpha  \eta +\eta ^2+16\right)^2}{24 \alpha ^4+20 \alpha ^3 \eta +2 \alpha ^2 \left(\eta
   ^2+40\right)-\alpha  \eta  \left(\eta ^2-32\right)-4 \eta ^2+64}\right)^{\half}, \\
  C_b &= \left( \frac{Q_m^2 \left(2 \alpha ^2+\alpha  \eta +4\right)}{L^4 V_0 \left(-2 \alpha  \eta -\eta ^2+8\right)}\right)^{\frac{1}{4}},
  \quad \quad
  K = \frac{4 (2 \alpha +\eta )}{(2 \alpha +\eta )^2+16}.
 \end{split}
\end{equation}
Note that this solution is identical to that in \eqref{earlyhyper}; here we have merely expressed all the solution
parameters explicitly in terms of those in the lagrangian.

As we already discussed in the introduction, the dilatonic scalar field in these solutions
drives the system towards strong coupling at the horizon\footnote{Recall that we are taking $\alpha K >0$.},
indicating a breakdown of the theory -- and in particular
denoting the failure of (\ref{SolRepeat}) to accurately describe the geometry.
Assuming such solutions can arise in a concrete string theory realization, as the coupling
$g = e^{-\a\phi}$
grows quantum corrections are expected to become important and to lead to a deformation of the geometry itself -- providing
an IR-completion of (\ref{SolRepeat}).
For the case of branes exhibiting Lifshitz scaling, this point was discussed in \cite{Goldstein:2009cv,Goldstein:2010aw}
and studied recently in \cite{Harrison:2012vy}.

Here we would like to follow the strategy of \cite{Harrison:2012vy} and add generic corrections to the gauge kinetic function,
meant to mimic the effects of adding quantum corrections in the theory. The analysis of corrections to the scalar potential
-- which we are assuming to be protected here -- would proceed in an entirely analogous manner, as will be clear shortly.
We will see that promoting
the gauge kinetic term to an expansion in powers of the coupling $g$,
\beq
\label{QMf}
f(\phi) = e^{2\a\phi} + \xi_1 + \xi_2 e^{-2\a\phi} + \ldots
\eq
will generate an attractor potential for the scalar field, allowing for
the existence of a minimum $\phi=\phi_H$ and in turn for $AdS_2 \times R^2$ solutions.
We should emphasize that -- although a simple expansion of the form of (\ref{QMf}), controlled by just two parameters $\xi_1,\xi_2$,
is enough to make our point -- we have also considered a generic $f(\phi)$ in our analysis.

Finally, we would like to note that in the recent study of dyonic solutions \cite{Kundu:2012jn} it has been shown that
(in certain regions of phase space, in which a small magnetic field perturbation is relevant in the infrared)
a similar $AdS_2 \times R^2$ IR completion is realized classically.
We refer the reader to \cite{Kundu:2012jn} for a discussion of the effects of adding a small magnetic field in the background
of an electric field, and of the behavior of the entanglement entropy in that context.

\subsection{$AdS_2 \times R^2$ as an exact solution} \label{ads2}

We are now ready to ask whether $AdS_2 \times R^2$ with a constant dilaton ($\phi=\phi_H$) is a solution to this system,
first classically, i.e. by setting $\xi_1=\xi_2=0$ in (\ref{QMf}),  and then with the inclusion of quantum corrections, by
allowing them to be non-zero. We will first assume that $f(\phi)$ is given by (\ref{QMf}), and then generalize it to an arbitrary function.
We start by taking the metric to be of the $AdS_2 \times R^2$  form,
\beq
\label{ads2r2metric}
ds^2 = L^2 \left( -r^2 dt^2 + \frac{dr^2}{r^2} + b_H^2 (dx^2 +dy^2) \right) \, ,
\eq
with $b_H$ a constant, and consider the two cases -- classical vs. quantum -- separately:

\begin{itemize}
\item
{\em Case (i): \; $\xi_1=\xi_2=0$}\\
We note first that (\ref{aEOM})  reduces to the simple condition
$\phi^{\, \prime}(r) = 0$, which clearly supports a
constant scalar $\phi=\phi_H$, independently of the form of $f(\phi)$ and $V(\phi)$.
The equation (\ref{cEOM}) for the scalar potential, which is also independent of $f$, is easily satisfied
\beq
\label{con0}
V_0 = \frac{e^{\, \eta\phi_H}}{L^2} \, ,
\eq
and guarantees that the overall sign of $V(\phi_H)$ is negative.
Plugging the expression for the potential in (\ref{bEOM}) then leads to
\beq
\label{con1}
e^{2\alpha\phi_H} = \frac{2b_H^4 L^2}{Q_m^2} \, ,
\eq
and finally (\ref{phiEOM}) gives the more interesting condition
\beq
\label{con2}
e^{2\alpha\phi_H} = - \frac{b_H^4 L^2}{Q_m^2} \frac{\eta}{\alpha} \, ,
\eq
which \emph{cannot} be satisfied if $\a$ and $\eta$ have the same sign.
Moreover, satisfying both (\ref{con1}) and (\ref{con2}) forces\footnote{In the double scaling limit of \cite{Hartnoll:2012wm},
i.e. $z\rightarrow \infty$, $\theta \rightarrow \infty$ with $\tilde{\eta} \equiv - \theta/z$ fixed,
our expression for $\beta$ takes the form $\beta = \frac{\tilde{\eta}}{(2\tilde{\eta}+2)}$.
Our condition $\beta=0$ is clearly solved only by $\tilde{\eta}=0$, which corresponds to $AdS_2 \times R^2$ \cite{Hartnoll:2012wm}.
Thus, the double scaling limit does not give rise to any additional solutions. We thank Sean Hartnoll for clarifying this point.}
\beq
\label{beta0}
\frac{\eta}{\alpha} = -2    \quad \Rightarrow \quad \beta = 0 \, .
\eq
Using (\ref{betagamma}), for finite $z$ this constraint translates into the following condition on the hyperscaling violating exponent,
\beq\label{th2}
\theta=2 \, ,
\eq
which is forbidden \cite{Huijse:2011ef} by the requirement\footnote{This relation is expected to hold for holographic duals of QFTs
that do not have large accidental degeneracies in their low energy spectrum.}
 (recall for us $d=2$)
\beq
\theta \leq d-1
\eq
that the entanglement entropy associated with the hyperscaling violating region obeys the area law (modulo
log corrections).
The constraint (\ref{beta0}) is also satisfied in the limit $z\rightarrow\infty$, as expected\footnote{In the
$z \rightarrow \infty$ limit Lifshitz metrics are known to reduce to $AdS_2 \times R^2$.}.
Here we will restrict our attention to the case of finite $z$ and $\theta <1$, with the latter condition meant to avoid
having to match onto scaling solutions with area-law violations.
Thus, just as in the Lifshitz case studied in \cite{Harrison:2012vy}, we see explicitly that \emph{classically} the theory
considered here --
with a constant magnetic field and an effective potential controlled by (\ref{fVform}) --
does not allow for an $AdS_2 \times R^2$ geometry in the IR -- apart from the two special cases
$\theta=2$ and $ z \rightarrow \infty$ discussed above.

\item
{\em Case (ii): \; $\xi_1,\xi_2 \neq 0$}\\
Next, we turn on the parameters $\xi_{1,2}$ which control our toy model for quantum corrections, with the expectation that they
will generate an effective potential for the scalar, stabilizing it at some constant value $\phi_H$.
It is easy to see that (\ref{aEOM}) and (\ref{cEOM}) are insensitive to the gauge kinetic function and therefore remain unchanged --
a constant scalar is still supported by the former, and the latter still reduces to (\ref{con0}).
The remaining two conditions, (\ref{con1}) and (\ref{con2}) respectively, are now modified and take the form
\bea
\label{con3}
e^{2\alpha\phi_H} + \xi_1 + \xi_2 e^{-2\alpha\phi_H}  &=& \frac{2b_H^4 L^2}{Q_m^2} \, , \\
\label{con4}
e^{2\alpha\phi_H}  - \xi_2 e^{-2\alpha\phi_H}  &=& - \frac{\eta}{\alpha}  \frac{ b_H^4 L^2}{Q_m^2} \, .
\ea
It is now clear that once $\xi_2$ is turned on it is possible to satisfy (\ref{con4}), provided that the condition
\beq
\label{xi2cond}
\left(1-\xi_2 \, e^{-4\a \phi_H} \right) < 0
\eq
is obeyed. The actual value $\phi_H$ at which the scalar is stabilized can be found by solving (\ref{con3}).
\end{itemize}

Clearly, this analysis can be easily redone with a more general form for the putative quantum corrections.
More precisely, parameterizing
\beq
f(\phi(r)) = e^{2\a\phi(r)} + \mathcal{G}(\phi(r)) \, ,
\eq
the conditions (\ref{con3}) and (\ref{con4}) are modified in the following way,
\bea
\label{con5}
e^{2\alpha\phi_H} + \mathcal{G}(\phi_H)  &=& \frac{2b_H^4 L^2}{Q_m^2} \, , \\
\label{con6}
2\alpha \, e^{2\alpha\phi_H}  + \partial_\phi \mathcal{G} (\phi_H) &=& - \eta  \frac{2 b_H^4 L^2}{Q_m^2} \, .
\ea
Thus, these relations provide \emph{constraints on the value and slope} of the arbitrary correction $\mathcal{G}(\phi)$ (evaluated
at the horizon) needed to obtain $AdS_2 \times R^2$ as a solution.
In particular, since only the first derivative of $\mathcal{G}$ affects the analysis, there is a certain amount of
`universality' in the structure of possible corrections.

Even though thus far we have taken $V(\phi)$ to be protected and left it untouched, corrections
to the potential can also be easily incorporated.
More specifically, letting $V(\phi) = - V_0 \, e^{-\eta\phi} + \mathcal{V}(\phi)$,
it's apparent from (\ref{cEOM})--(\ref{bEOM}) that (\ref{con5}) will remain the same, while (\ref{con6}) will be modified to
\beq
2\alpha \, e^{2\alpha\phi_H}  + \partial_\phi \mathcal{G} (\phi_H) = - \frac{2 b_H^4 L^2}{Q_m^2}
\Bigl[\eta +  \eta  L^2 \mathcal{V}(\phi_H) + L^2 \partial_\phi \mathcal{V}(\phi_H) \Bigr] \, ,
\eq
with the correction $\mathcal{V}$ satisfying the condition $\mathcal{V}(\phi_H) = V_0 e^{-\eta \phi_H} - \frac{1}{L^2}$.
Although here we have just sketched the analysis, the simple point we would like to stress
is that the emergence of an $AdS_2$ factor in the infrared is in no way restricted to the specific choice (\ref{QMf}),
but is in fact much more robust.
Clearly these types of conditions (restricting the value and slope of $\mathcal{G}$ and $\mathcal{V}$ at the horizon) apply to arbitrary corrections,
independently of whether their origin is quantum mechanical or classical.
In particular, they illustrate the emergence of $AdS_2 \times R^2$ at the classical level in the setup of \cite{Kundu:2012jn},
where the presence of electric and magnetic fields gives rise to a trapping potential for the dilatonic scalar,
in appropriate regions of phase space.

\subsection{Perturbations about $AdS_2 \times R^2$ }\label{per}

Having constructed and established conditions for the existence of an $AdS_2 \times R^2$ solution to the system \eqref{GeneralLag},
we proceed to construct solutions that evolve from $AdS_2 \times R^2$ in the IR to $AdS_4$ in the UV.
In particular, among solutions which interpolate between the two fixed points,
 we wish to explore the possibility of the existence of an intermediate geometry characterized by non-trivial scaling exponents
$z$ and $\theta$. In order to achieve our goal, we begin by classifying all linear perturbations to the $AdS_2 \times R^2$
geometry that are \emph{irrelevant} in the IR.

Borrowing notation from \cite{Harrison:2012vy},
we perturb around the infrared $AdS_2 \times R^2$ solution obtained in \S \ref{ads2} with the ansatz
\begin{equation}\label{ads2per}
 a(r) = r(1 + d_1 r^\nu) \, , \quad
 b_{r} = b_{H} (1 + d_2 r^\nu)\, , \quad
 \phi(r) = \phi_{H} (1 + d_3 r^\nu) \, ,
\end{equation}
where the magnitude of $d_1, ~d_2$ and $d_3$ is proportional to the amplitude of the perturbation and is
assumed to be small (we work at leading order in these parameters).
Einstein's equations lead to the conditions
\begin{equation}
 \begin{split}
  d_2 (-1 + \nu^2) &= 0,\\
  d_2 (-1 + \nu) &=0,\\
  d_2 (2+\nu+\nu^2) + d_1 (2+3 \nu + \nu^2 )+ d_3 \eta \phi_{H} &=0 \, ,
 \end{split}
\end{equation}
while the scalar equation gives rise to the following constraint at leading order in the perturbations,
\begin{equation}
\begin{split}
 d_3 \phi_{H} &\left(2 e^{2 \alpha  \phi_{H}} \left(-4 \alpha ^2+\eta ^2 + 4 \nu  (\nu +1)\right)+\xi_1 \left(-2 \alpha  \eta
   +\eta ^2+4 \nu  (\nu +1)\right)\right)
\\  & \qquad \qquad \qquad \qquad \qquad \qquad \qquad \quad= 4 \eta  d_2 \left(2 e^{2 \alpha \phi_{H}} + \xi_1 \right)\, .
\end{split}
\end{equation}
Note that the apparent absence of $\xi_2$ in the above conditions is due to the fact that we have eliminated it using
\eqref{con3}, as it was more convenient than solving for $\phi_{H}$.

Clearly, to ensure that the modes are indeed irrelevant in the IR we need $\nu > 0$ to hold -- this guarantees that they become more and more
unimportant as $r$ decreases.
There are two sets of solutions to these conditions which appear as modes that are irrelevant in the IR:

{\it Mode1:} The first mode corresponds to the following solution
\begin{equation}
 \begin{split}
  \nu^{(1)} &= 1, \\
  d_2^{\, (1)} &= -\frac{3 d_1^{(1)} \left(2 \left(4 \alpha ^2-\eta ^2-8\right) e^{2 \alpha  \phi_{H}} +\xi_1 \left(2 \alpha  \eta -\eta
   ^2-8\right)\right)}{4 \left(\left(4 \alpha ^2-2 \left(\eta ^2+4\right)\right) e^{2 \alpha  \phi_{H}}+\xi_1 \left(\alpha  \eta -\eta
   ^2-4\right)\right)}, \\
  d_3^{\,(1)} &= -\frac{3 \eta  d_1^{(1)} \left(2 e^{2 \alpha  \phi_{H}}+\xi_1\right)}{\phi_{H} \left(\left(-4 \alpha ^2+2 \eta ^2+8\right) e^{2
   \alpha  \phi_{H}}+\xi_1 \left(-\alpha  \eta +\eta ^2+4\right)\right)} \, .
 \end{split}
\end{equation}
Note that in this solution $d_1^{\,(1)}$ is arbitrary, and its value
sets the amplitude of the perturbation.

{\it Mode2:} For the second mode, the solution is given by
\begin{equation}
 \begin{split}
  \nu^{(2)} &= \frac{\sqrt{\left(2 e^{2 \alpha  \phi_{H}}+\xi_1 \right) \left(\left(8 \alpha ^2-2 \eta ^2+2\right) e^{2 \alpha  \phi_{H}}+\eta
   \xi_1 (2 \alpha -\eta )+\xi_1\right)}-2 e^{2 \alpha  \phi_{H}}-\xi_1}{2 \left(2 e^{2 \alpha  \phi_{H}}+\xi_1\right)}, \\
  d_1^{\, (2)} &= \frac{\text{A}}{\text{B}} \, , \quad \quad  d_2^{\,(2)} = 0 \, , \\
  \text{A} &\equiv 4 \eta  \, d_3^{\,(2)} \, \phi_{H} \left(2 e^{2 \alpha  \phi_{H}}+\xi_1\right) \, ,\\
  \text{B} &\equiv\left(-4 \left(\sqrt{\left(2 e^{2 \alpha  \phi_{H}}
+\xi_1\right) \left(\left(8 \alpha ^2-2 \eta ^2+2\right) e^{2 \alpha  \phi_{H}}+\eta  \xi_1 (2 \alpha -\eta
   )+\xi_1\right)}+\xi_1\right) \right. \\ &\left.+\left(-8 \alpha ^2+2 \eta ^2-8\right) e^{2 \alpha  \phi_{H}}+\eta  \xi_1 (\eta -2
   \alpha )\right) \, .
 \end{split}
\end{equation}
Here $d_3^{\, (2)}$ is the parameter whose value determines the amplitude of the perturbation.

\subsection{The asymptotic UV region}
\label{UVa}

Before proceeding to construct the numerical solutions, we pause briefly to understand the asymptotic UV behavior of our system.
As mentioned previously in \S\ref{intro}, we are particularly interested in solutions which asymptote to $AdS_4$ in the UV,
in order to be able to apply standard holographic interpretations to our bulk physics.
However, the scalar field potential \eqref{fVform} that we have considered thus far does not admit
a minimum in the UV corresponding to a negative cosmological constant, and therefore does not support $AdS_4$ asymptotically.
Recall that in this note we are allowing for a constant magnetic flux only. In the ultraviolet, the effect of its gauge kinetic term becomes
negligible compared to that of the scalar field potential, so that the effective potential of the system  is controlled entirely by $V(\phi)$.
Thus, in the far UV region the scalar asymptotes to the extremum of the scalar potential, which in the case of \eqref{fVform} is zero\footnote{The
scalar approaches $\pm \infty$ in the UV, with the sign determined by the sign of $\eta$.}.

This problem can be easily addressed by appropriately modifying the potential (as discussed e.g. in \cite{Huijse:2011ef}),
and in particular in such a way not to affect the qualitative behavior of the system in the infrared.
As an example, one can take the potential to be of the simple form $V(\phi) \propto - V_0 \cosh\eta\phi$,
with a (negative) minimum at $\phi =0$, or more generally by choosing it to be of the form
\beq
\label{newpot}
V(\phi) = - V_0 \left(  e^{-\eta \phi} + c_1 e^{\eta_1 \phi}  \right) \, ,
\eq
which allows for a minimum at a non-zero value of the scalar,
\beq
\label{phiUV}
\phi_{uv} = \frac{1}{\eta+\eta_1} \ln \left(\frac{\eta}{c_1 \eta_1} \right) \, .
\eq
It's then easy to see that the scalar potential at the minimum takes the value
$V(\phi_{uv}) = - V_0 e^{-\eta \phi_{uv}} \left( 1+ \frac{\eta}{\eta_1}\right)$,
corresponding to a negative cosmological constant as long as the quantity in parenthesis is positive.
For simplicity from now on we will assume that $\eta_1=\eta$.

Note that a potential like that of (\ref{newpot}) can in principle (for appropriate parameter choices)
induce a near-horizon $AdS_2 \times R^2$ region without the need for modifications to the gauge kinetic function (recall
our discussion at the end of \S \ref{ads2}).
In the simple $\eta=\eta_1$ case we are discussing here, the existence of $AdS_2 \times R^2$ in the infrared
-- assuming that $f(\phi) = e^{2\alpha\phi}$ is left unchanged -- leads to the following
expression for the near-horizon value of the scalar,
\beq
e^{2 \eta \phi_H} = \frac{\eta+2\alpha}{c_1 (\eta-2\alpha)} \, .
\eq
Notice that this condition cannot be met e.g. when $c_1 ( \eta - 2\alpha) < 0$ and $\eta+\alpha>0$.
Thus, as long as $\eta$ and $c_1$ are chosen to lie within the range above, the potential alone will not be enough to
generate an infrared $AdS_2$ region -- the types of (quantum) corrections to the gauge kinetic function we have introduced
will still be needed, and the qualitative near-horizon behavior we have discussed will
remain unaffected by the modifications to the original scalar potential.

Since to describe the entire flow from the IR to the UV we will resort to numerics, for the rest of the discussion
in this subsection we will work with\footnote{These parameters satisfy the conditions
$c_1 ( \eta - 2\alpha) < 0$ and $\eta+\alpha>0$ discussed above.}
\begin{equation}\label{numval}
 \alpha=\sqrt{3},\ \xi_1 =0,\  \xi_2 =1,\  V_0 = 3 \times 10^4,\  c_1 = 10^{-4}, \ \eta_1 = \eta = \frac{2}{\sqrt{3}}.
\end{equation}
Note that these values (otherwise unmotivated) have been chosen to obtain \emph{large}
regions of hyperscaling violation in the numerical plots that we present in \S \ref{num}.
With a more extensive numerical analysis we expect to find similar flows for a broad region of parameter space, and in particular
for more `natural' values\footnote{Note that for these parameters, the value of the scalar at the UV minimum of the potential is
$\phi_{uv}= \frac{1}{2\eta}\ln \left( \frac{1}{c_1} \right) $. Thus, increasing $c_1$ corresponds to lowering the value of $\phi_{uv}$.}
of $\{ V_0, c_1\}$.
However, we recall that our main interest in this note is to probe the IR behavior of solutions with hyperscaling violation,
and in particular to \emph{identify cases} in which near the horizon they are replaced by $AdS_2 \times R^2$.
As a result, here we will content ourselves with presenting an explicit realization of the flow we are after,
without performing a more exhaustive numerical analysis.

We can easily verify that for the parameter choice \eqref{numval}
we continue to have an infrared $AdS_2 \times R^2$ solution with two irrelevant
perturbations, as described in \S \ref{ads2} and \S \ref{per}.
The corresponding $AdS_2 \times R^2$ parameters are then given by
\begin{equation}
 \phi_H = -0.1, \quad b_H = 13.74 \sqrt{Q_m}, \quad L = 0.0054,
\end{equation}
with the irrelevant fluctuations about this geometry -- of the form \eqref{ads2per} -- given by
\begin{equation}\label{paraset}
\begin{split}
 &\text{{\it mode 1}}: \nu^{(1)}= 1,\quad d_2^{\, (1)} = - 2.99 \ d_1^{\,(1)}  ,\quad d_3^{\,(1)} = 5.19 \ d_1^{\,(1)} \, ,\\
 &\text{{\it mode 2}}: \nu^{(2)}= 1.21,\quad d_1^{\,(2)} = 0  ,\quad d_2^{\,(2)} = -0.16 \ d_3^{\,(2)} \, .\\
\end{split}
\end{equation}
In the extreme UV the scalar field settles to the minimum of the (effective) potential \eqref{newpot}, which in this case occurs when
\begin{equation}
 \phi_{uv} = 3.99.
\end{equation}
The value of the potential evaluated at this minimum then provides the negative cosmological constant needed to support
the asymptotic $AdS_4$ geometry,
\begin{equation}
\begin{split}
 ds^2 = L^2 \left(-(A r)^2 dt^2 + \frac{1}{(A r)^2} dr^2 + B^2 r^2 \left( dx^2 + dy^2\right) \right)\\
\end{split}
\end{equation}
where $A$ and $B$ are numbers fixed by the choice of parameters in the IR.

Next, we would like to discuss briefly linear fluctuations about the $AdS_4$ UV geometry,
\begin{equation}
 a(r) = A r (1+ \epsilon \lambda_1 r^\nu ) \, ,  \quad b(r) = B r (1 + \epsilon \lambda_2 r^\nu ) \, , \quad \phi(r) = \phi_{uv}
 + \epsilon \lambda_3 r^\nu \, ,
\end{equation}
where we emphasize that the leading order value $\phi_{uv}$ of the scalar field is determined by minimizing the potential \eqref{newpot}.
If we want to satisfy the equations of motions up to linear order in $\epsilon$  we are forced to choose
\begin{equation}
\label{uvperts}
 \lambda_1 =0, \quad \lambda_2=0, \quad \nu = -2, -1 \, ,
\end{equation}
where we have taken into account the fact that
 in the far UV the magnetic flux contribution to the effective potential is suppressed compared to the remaining terms.
Thus, from (\ref{uvperts}) we conclude that the scalar field approches its UV value
(the minimum of the potential) either as $r^{-2}$ or as $r^{-1}$.

Finally,
we note that the hyperscaling violating geometry constructed in \eqref{SolRepeat}, \eqref{SolRepeat2}
and \eqref{SolRepeat3} ceases to be
an exact solution with the modified potential \eqref{newpot}, just like it ceases to be an exact solution in the presence
of the corrections \eqref{QMf} to the gauge kinetic term.
However, even in the presence of these modifications,
a hyperscaling violating geometry can be realized in an intermediate
region, where the effects of such terms are negligible.
For the choice of parameters in \eqref{numval}, this hyperscaling violating geometry has $\theta = -2$ and $z=3/2$.
We shall now proceed to construct numerical solutions realizing the type of flow we have discussed,
admitting a regime of hyperscaling violation.

\section{Numerical solution}
\label{num}

In this section we construct numerical solutions to the set of equations \eqref{aEOM}-\eqref{phiEOM}
which flow from $AdS_2 \times R^2$ in the deep IR to an intermediate region displaying both
hyperscaling violation and Lifshitz-like scaling.
As we have discussed in \S\ref{s:ads2}, there is a two parameter set of
irrelevant deformations to this infrared $AdS_2 \times R^2$ geometry -- here we will follow
these deformations numerically (for specific parameter choices) as they evolve towards the UV.

We will consider first the original single-exponential potential \eqref{fVform} which does not support
an asymptotic $AdS_4$, as discussed in \S \ref{UVa}.
In this case we will see that -- by fine-tuning sufficiently the deformation parameters --
it is possible to obtain an hyperscaling violating geometry in the UV\footnote{While performing
the numerical analysis we found that the hyperscaling violating geometry broke down at
some point in the extreme UV. However, this point could be pushed further and further away with better
accuracy of the fine-tuning. This leads us to conclude that
such a break down is essentially a numerical artifact.}.
After studying numerically several examples,
we suspect that in the two parameter set of deformations there is a line along which the hyperscaling
violating solution exists as the UV geometry. We present our numerical plots for this case in
\S \ref{asymHV}.

We will then go on to consider a potential of the form of \eqref{newpot}, which admits a minimum in the ultraviolet
corresponding to a negative cosmological constant.
In this case, starting from $AdS_2 \times R^2$ in the IR, we numerically shoot to obtain an $AdS_4$
geometry in the UV.
Again, for sufficient fine-tuning of the deformation parameters
one can pass through a regime of hyperscaling violation, keeping
the UV $AdS_4$ geometry intact.
In the intermediate hyperscaling violating region the dilaton
decreases logarithmically as we move towards the IR, and the geometry transits into $AdS_2 \times R^2$
when the scalar reaches $\phi_H$ -- this happens when the toy quantum correction terms (controlled by
$\xi_1$ and $\xi_2$) become important.
On the other hand, as the dilaton moves away
from the hyperscaling violating geometry and approaches the UV, it settles to the minimum of the (effective) potential, where the geometry is $AdS_4$.
We present out numerical plots for this case in \S \ref{asymads}.

\subsection{Asymptotically hyperscaling violating geometry}\label{asymHV}
For performing our numerical analysis with the potential \eqref{fVform},
we have chosen the following set of values for the lagrangian parameters
\begin{equation}
 \alpha =1, \quad V_0 =1, \quad \xi_1 =1, \quad \xi_2 = 0.5 \, ,
\end{equation}
and have taken the value of the constant magnetic flux to be
\begin{equation}
 Q_m = 2 \, .
\end{equation}
For this set of parameters, we have obtained the optimal values for the amplitude of the IR irrelevant fluctuations
which lead to hyperscaling violation in the UV. In this note, we present the results for two sets of parameters for this
choice of potential
\bea
\label{paraset}
 \text{{\it set 1}}&:& \eta = 0.1\, , \quad d_1^{(1)} =-0.001\, , \quad d_3^{(2)}=-0.141202\, ,
\quad \gamma =0.02 \, , \quad \beta =0.21\, , \nn \\
 &&  \theta =-0.21 \, , \quad z =4.9 \, , \nn \\
 \text{{\it set 2}}&:& \eta = -0.1\, , \quad d_1^{(1)} =-0.001\, , \quad  d_3^{(2)}=-0.239086 \, ,
 \quad \gamma =-0.02 \, , \beta =0.18\, , \nn \\
 &&  \theta =0.19 \, , \quad z =5.1 \, .
\ea
The numerical plots\footnote{As a test of our numerics we have reproduced the plots
in  \cite{Harrison:2012vy}  for $\eta=0$ and $d_1^{(1)} =-0.001, d_3^{(2)}= -0.1779$.}
of our solutions for these parameter choices are shown in Fig. \ref{fig:funcplots1} and Fig. \ref{fig:funcplots2}.
We have tested that for the chosen range of parameters the null energy condition holds (see appendix \ref{app:nullcond}),
indicating that we have reasonable matter and valid gravitational solutions in the classical regime, away from the deep IR.
For the first set ({\it set1}) of parameters the hyperscaling violating
coefficient $\theta$ is negative (see also \cite{Dong:2012se} for explicit string theory realizations of systems with hyperscaling
violation with $\theta<0$ and $z=1$). In {\it set2} we have chosen parameters so as to obtain $\theta>0$.

\FIGURE{
\centering
\includegraphics[width=\textwidth]{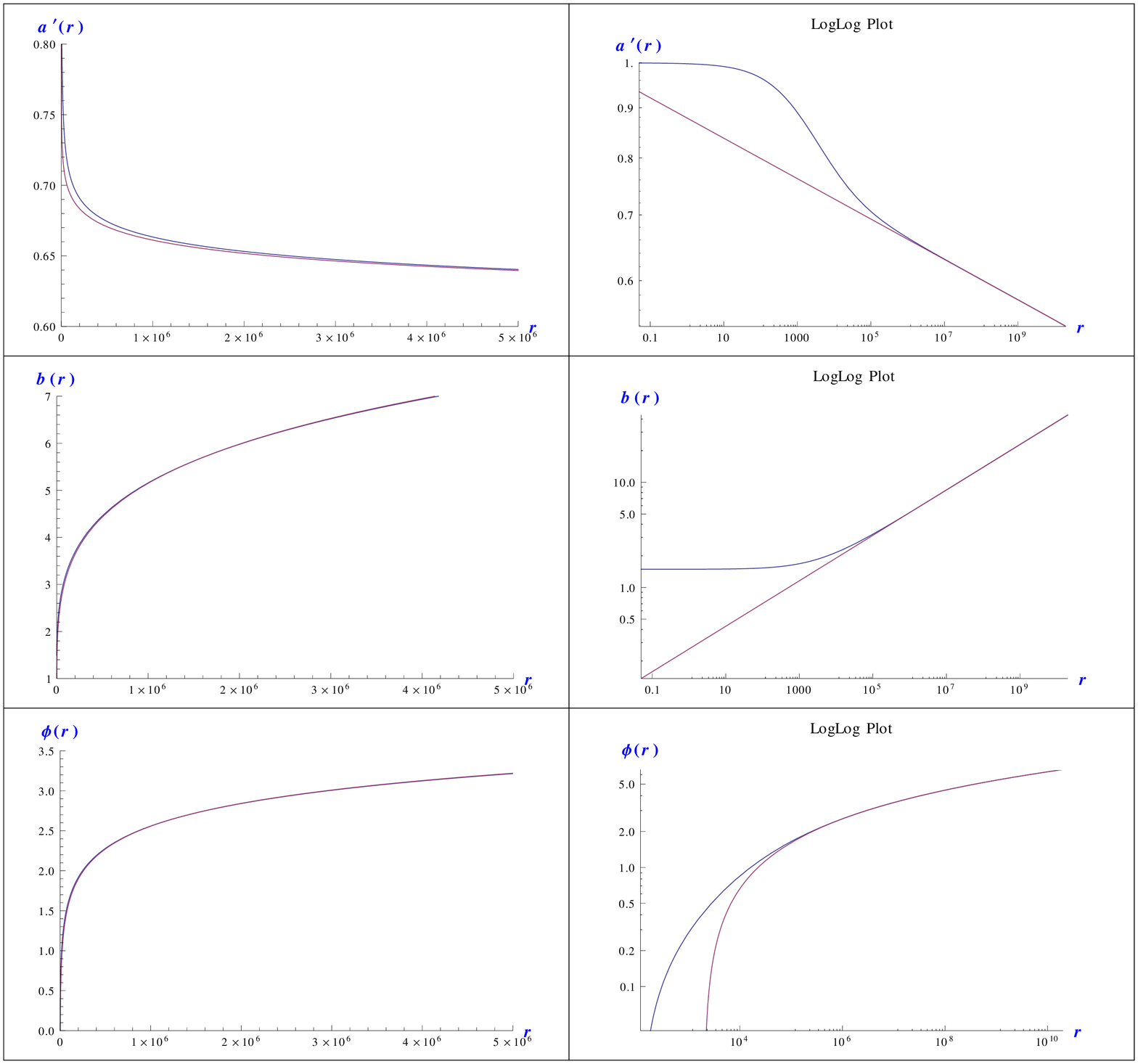}
\caption{Plots of the metric functions and dilaton for the parameter set 1 ($\eta$ = 0.1).
Note that the red line represents a hyperscaling violating and Lifshitz-like scaling solution with  $\{\theta=-0.21, z=4.9\}$, while the
blue line represents our numerical solution. The fact that a hyperscaling violating regime emerges in the UV  is
clear from the matching of the two plots in that region.}
\label{fig:funcplots1}
}

\FIGURE{
\centering
\includegraphics[width=\textwidth]{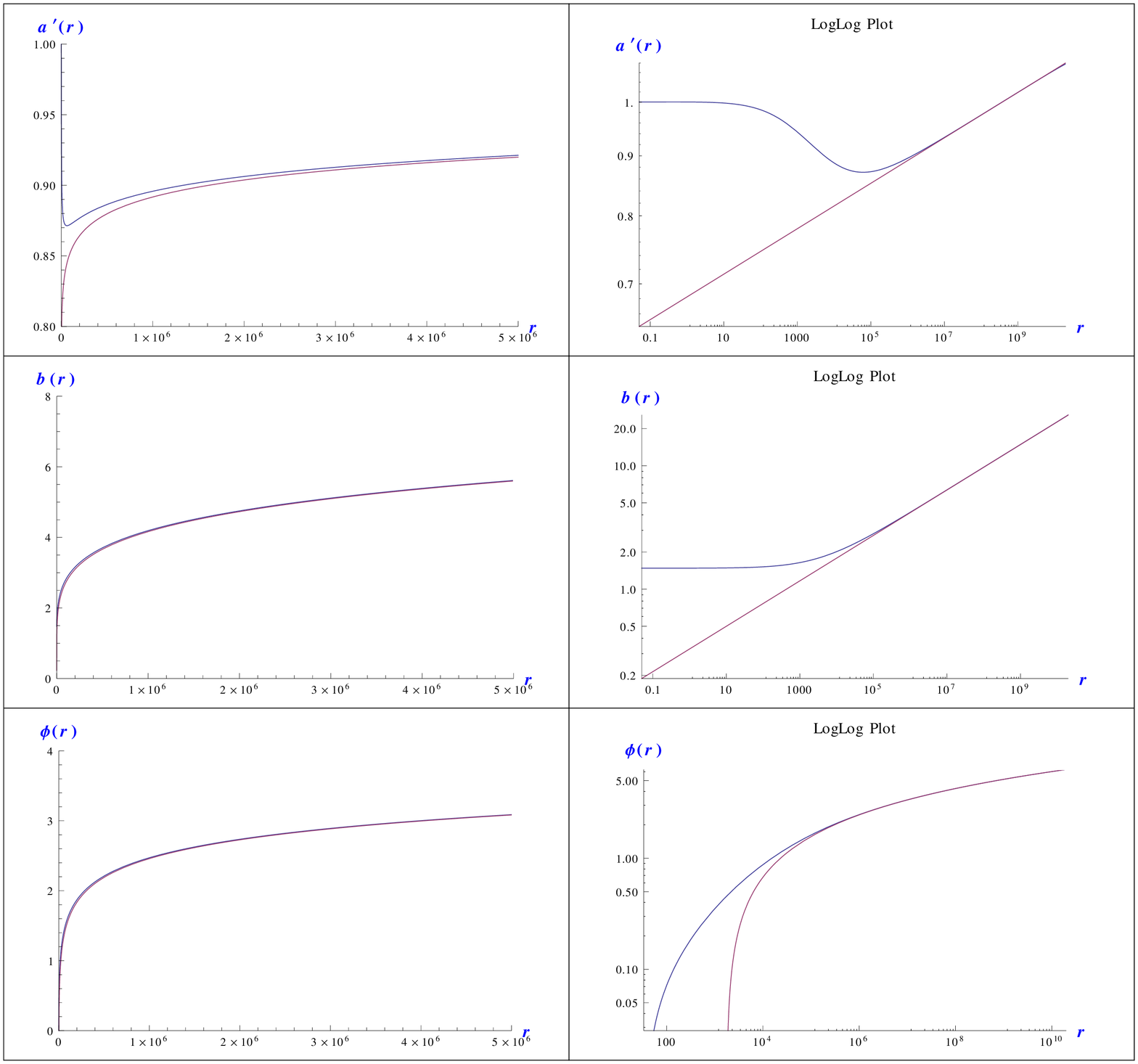}
\caption{Plots of the metric functions and dilaton for the parameter set 2 ($\eta$ = -0.1).
As in the previous figure, the red line represents a hyperscaling violating and Lifshitz-like scaling solution with $\{\theta=0.19, z=5.1\}$, while the
blue line represents our numerical solution.
The fact that the hyperscaling violating solution emerges in the UV is
clear from the matching of the two plots in that region.}
\label{fig:funcplots2}
}

In all the three functions in Fig. \ref{fig:funcplots1} and Fig. \ref{fig:funcplots2}, we see a distinct scaling region in the UV.
From the plots on the left hand side it is apparent that beyond a certain point in the radial direction, there is precise
agreement between the numerical solution (the blue line) and the corresponding
hyperscaling violating solution (the red line) with the same scaling exponents.

The presence of the scaling region in the UV is even more apparent from the log-log plots on the
right hand side of Figures \ref{fig:funcplots1} and \ref{fig:funcplots2}.
Deep in the IR, we start with the $AdS_2 \times R^2$ solution, for which $a^{\,\prime}(r)$ and
$b(r)$ are constants, as clearly visible in their respective plots. The scalar field in this region assumes
a constant value determined by \eqref{con3}.
As we move away from the IR we enter the hyperscaling violating regime, where
the functions $a^{\,\prime}(r)$ and $b(r)$ scale with the exponents $\{-0.02 ,0.21 \}$ and $\{0.02 ,0.18\}$ respectively, for the chosen set of
parameters, and the scalar field runs logarithmically. As visible from the plots,
in all of the three functions in Figures \ref{fig:funcplots1} and \ref{fig:funcplots2}
the matching with the hyperscaling violating solutions occurs at the same radial distance,
confirming the fact that a hyperscaling violating region does indeed emerge.

Before closing this section, we emphasize once again that hyperscaling violation in the UV arises for
extremely fine-tuned values of the parameters $d_1^{(1)}$ and $d_3^{(2)}$.
When the deformation parameters are away from these fine-tuned values
we are generally led to singular geometries in the UV.
In certain cases, however,
one finds that an $AdS_4$ geometry emerges beyond the hyperscaling violating region, where the scalar field
runs to $\pm \infty$ (towards the extremum of the unmodified potential).
Finally, it may also be possible to obtain other scaling geometries in the UV, but a conclusive statement along this direction can
only be made with a more detailed exploration of the parameter space of deformations\footnote{We would like to thank Blaise Gout\'{e}raux for
pointing out that these may correspond to neutral scaling solutions, which have non-zero $\theta$ but $z=1$.}.

\subsection{Asymptotically $AdS_4$ geometry}\label{asymads}

In this subsection we present numerical plots for the case corresponding to the modified potential \eqref{newpot},
which we recall was constructed to support $AdS_4$ in the UV.
\FIGURE{
\centering
\includegraphics[width=\textwidth]{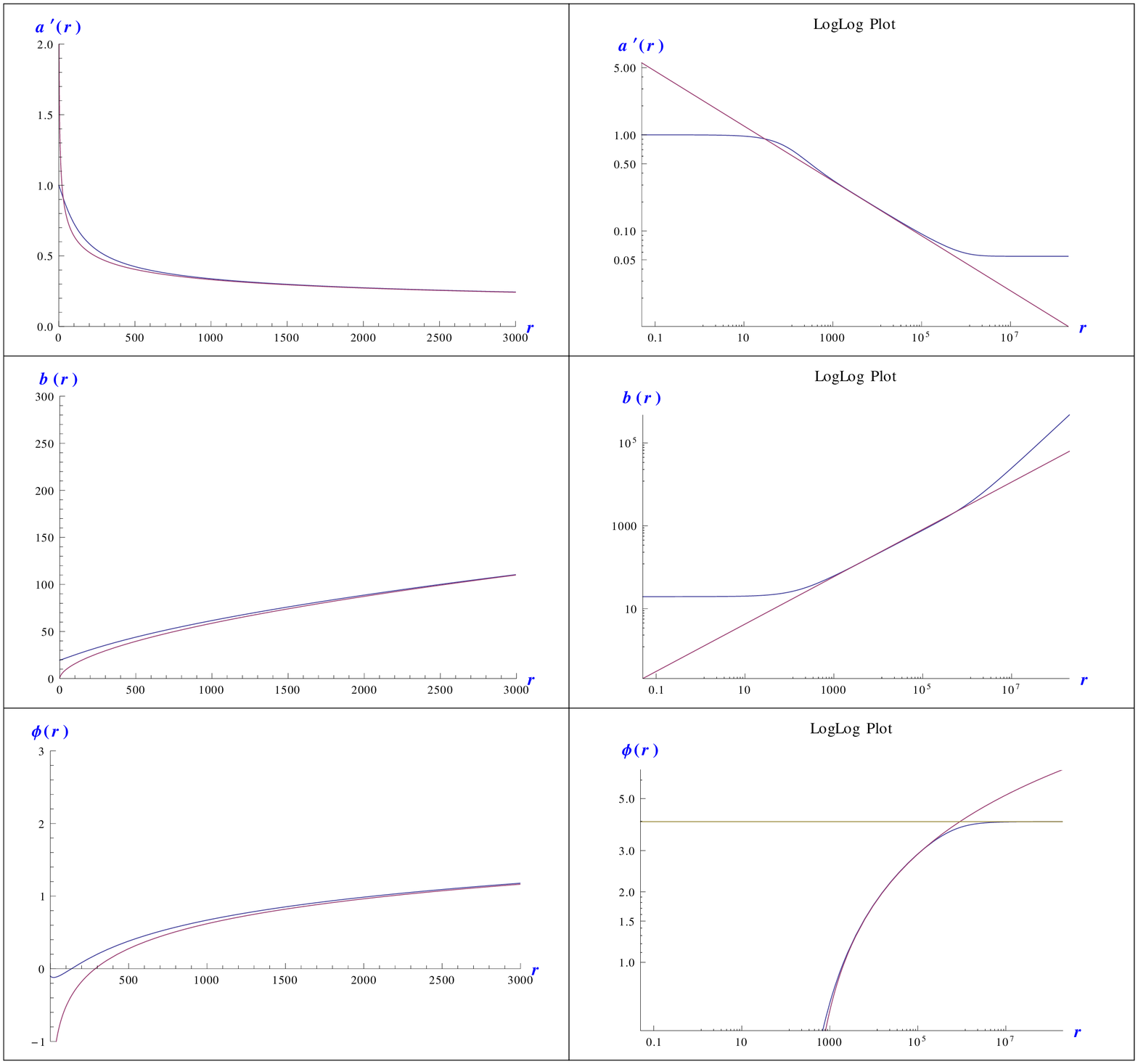}
\caption{Plots of the metric functions and dilaton with the modified potential. Here the red lines represent a hyperscaling violating solution
(with $\theta =-2$ and $z=3/2$) while the
blue lines represents our numerical solution.
The fact that the hyperscaling violating solution emerges in the intermediate region is
clear from the matching of the two plots there. In the log-log plot for $\phi$, the yellow line is the constant UV value $\phi_{uv}=3.99$.
The numerical plot (blue line) approaches the yellow line like $\phi_{uv} - 2\times 10^5 \ r^{-1}$, as expected.}
\label{fig:funcplots3}
}
We make the same choices as in \eqref{numval}, and take the remaining parameters to be
\begin{equation}\label{numval2}
 Q_m = 2 \, , \quad d_1^{(1)} =-0.001 \, , \quad d_3^{(2)}= 0.00215 \, .
\end{equation}
The numerical plots for this case are shown in Fig. \ref{fig:funcplots3}.

We begin with $AdS_2 \times R^2$ in the IR, followed by an intermediate scaling region with $\theta =-2$ and $z=3/2$.
The geometry ultimately goes over to $AdS_4$ in the UV, which is indicated by the fact that
$a'(r)$ is a constant while $b(r)$ grows linearly with $r$.
In the ultraviolet, the dilaton settles to $\phi_{uv} = 3.99$, the value which minimizes the scalar
potential \eqref{newpot}, as can be read off from \eqref{phiUV}. In particular, the dilaton approaches its UV value
scaling as $r^{-1}$,  which is consistent with the asymptotic linear fluctuation analysis sketched in \S \ref{UVa}.
The $r^{-1}$ fall-off behavior is best seen in the log-log plot at the bottom of Fig. \ref{fig:funcplots3}.

In closing we would like to emphasize that, just like in the previous cases,
the value of $d_3^{(2)}$ has to be fine-tuned to a very high accuracy in order
to obtain the intermediate hyperscaling violating regime.
It is also interesting to note that there is a nearby point
(in the two-dimensional phase space of IR irrelevant modes)
on approaching which we obtain larger and larger intermediate regions of hyperscaling violation.
Right beyond this point the solutions diverge in the UV.
This makes us wonder if the following picture is true. There may
exist a subspace of this two dimensional phase space for which we have $AdS_4$ asymptotically --  and
it is only when we approach the boundary of this subspace that the intermediate hyperscaling violating regime emerges.
Confirming this picture would require a more detailed scan of the parameter space.
However, an extensive numerical study is beyond the scope of this note.

\section{Discussion}\label{disc}

Einstein-Maxwell-dilaton theories with simple scalar field profiles have proven to be a rich playground for
generating solutions with interesting scaling properties.
In particular, they have been shown to give rise to holographic realizations of
condensed matter systems characterized by both Lifshitz scaling and hyperscaling violation -- parametrized, respectively,
by the exponents $z$ and $\theta$.
Solutions realizing such scalings are of particular interest because -- for appropriate regions of
parameter space -- they give rise to phases which violate the area law for entanglement entropy.

In this note we have examined a class of Einstein-Maxwell-dilaton theories that admits magnetically
charged solutions with non-trivial Lifshitz-like scaling and hyperscaling violation.
These solutions are well-known to be supported by a logarithmically running scalar,
which drives the system towards strong coupling near the horizon
-- thus, the solutions are not `IR-complete,' and are expected to be modified by quantum corrections no
longer negligible in the strongly coupled region.
Our main interest here was precisely to understand the fate of these hyperscaling violating geometries as the theory is pushed towards the IR.
By taking into account
the generic features expected from quantum effects,
we have argued that the deformed theory
 admits a new class of exact solutions with an $AdS_2 \times R^2$ geometry emerging in the deep IR. The latter ceases to be a solution
 if the (quantum) deformation parameters are taken to zero.
We find that the only two exceptions -- i.e. situations in which the IR-completion occurs classically --
are the special cases with $z\rightarrow \infty$ (for which the geometry is known to reduce to
 $AdS_2\times R^2$), and  $\theta=2$ for finite $z$. The latter, however, is associated with a violation of the area law of entanglement entropy
 in the regime of validity of the hyperscaling violating geometry.

We have started our analysis by deriving a set of constraints on the form that a generic gauge kinetic function and scalar potential
would need to have in order to engineer hyperscaling violating solutions, recovering the simple system
\beq
\label{DiscLag}
{\mathcal L} = R-2 (\partial \phi)^2
- e^{2\alpha\phi} F^2
- V_0 e^{-\eta\phi} \, ,
\eq
with the parameters $\alpha$ and $\eta$ dictating the structure of the scaling exponents $z$ and $\theta$.
Having done that, we have deformed the (classical) theory by including corrections to the gauge kinetic function,
which one can parametrize as an expansion in powers of the coupling $g=e^{-\alpha\phi}$,
\beq
\label{fExpCon}
f(\phi) \rightarrow e^{2\alpha\phi} + \xi_1 + \xi_2 e^{-2\alpha\phi} + \ldots
\eq
Although keeping the first two terms in the expansion is enough to stabilize the dilaton
-- and to generate $AdS_2 \times R^2$ --
we have performed the analysis with \emph{arbitrary} corrections to the gauge kinetic function.
However, since only $f$ and its first derivative $\partial_\phi f$ affect the analysis,
cutting off a generic expansion of the type of (\ref{fExpCon}) does not qualitatively change the result --
i.e., we see a form of `universality' in the structure of the conditions for the existence of $AdS_2 \times R^2$.
We should also emphasize that keeping the form of the correction generic makes it feasible to include in a straightforward
way possible corrections to the scalar potential, which would play an analogous role to those discussed here.
In particular, this may prove useful for realizing these geometries within concrete string theory embeddings, and
relating to known supergravity solutions (see e.g. \cite{Barisch:2011ui,Donos:2011pn}).

Finally, we constructed numerical solutions to the quantum-corrected action which interpolate
between two fixed points, $AdS_4$ in the UV and $AdS_2 \times R^2$ in the IR -- in the presence of a constant magnetic flux.
The most novel feature of this interpolating solution is the emergence of an intermediate region with hyperscaling and Lorentz
violation, which is precisely what we were after\footnote{ In particular, the intermediate scaling region was obtained for
parameter choices that correspond to $\theta <d-1$, i.e. not violating the area law of entanglement entropy.}.
This realizes concretely the intuitive picture that the scaling solutions are not generically expected to survive in the deep IR --
where the low-energy breaks down -- but should be modified appropriately once quantum effects are taken into account.
Precisely the same type of flow was already seen in \cite{Harrison:2012vy}
for the Lifshitz case.
Although our focus here has been on magnetically charged branes, the electrically charged case is equally interesting.
In that context, however, the dilatonic scalar drives the system to weak coupling, and $\alpha^\prime$ corrections are
believed to become important.

Moreover, some of these hyperscaling violating solutions can be IR-completed already at the classical level,
by turning on, in addition to an electric field,
a \emph{small} magnetic field \cite{Kundu:2012jn}. In this type of dyonic system
the effective potential is such as to stabilize
the dilatonic scalar at the horizon, giving rise to an $AdS_2 \times R^2$ description
-- this occurs for regions of parameter space in which the magnetic field corresponds to a relevant perturbation in the IR.
Thus, we should emphasize that our analysis in this note complements\footnote{The range
for the lagrangian parameters ${\alpha,\eta}$ in this note corresponds to Case II of \cite{Kundu:2012jn}.}
 that of \cite{Kundu:2012jn}, which stresses that turning on even a small amount
of magnetic field can have a dramatic effect on the behavior of the system -- and in particular on that of the entanglement entropy.

In summary,
our construction in this note adds to the large landscape of vacua that may find interesting applications to
condensed matter systems.
The emergence of the $AdS_2$ factor in the deep IR feeds into the
well-known puzzle associated with the extensive ground state entropy of the extremal Reissner-Nordstrom $AdS_2 \times R^2$ region,
and ties into the question of what is the true ground state of these theories (see \cite{Donos:2011qt} for a discussion of the
(in)stability of magnetically charged $AdS_2 \times R^2$ backgrounds).
Moreover, one of the elusive goals in the study of hyperscaling violating theories has been to find string theory embeddings of these solutions
for general $\theta$ and $z$. It would be interesting to lift our toy model to the framework of string theoretic constructions.
Along these lines, solutions such as the ones obtained in this note also exists for classical
actions with more general dilaton potentials
(see e.g. \cite{Donos:2011pn}, which also contains magnetic solutions interpolating
between $AdS_2 \times R^2$ in the IR and $AdS_4$ in the UV).
Although hyperscaling violating solutions may not be exact solutions in such systems
(as indicated by the analysis in \S\ref{enghyper}), they do appear
as intermediate geometries in these set up. In fact this could be the reason why
they have not been easily observed in the study of such general systems within the framework of supergravity.
We leave further study of this question to future work.

\vskip1cm
\noindent
{\bf \large{Acknowledgement}}\\
We would like to thank Nabamita Banerjee, Jerome Gauntlett, Blaise Gout\'{e}raux, Sean Hartnoll, Rene Meyer, Sandip Trivedi and Scott Watson for useful discussions.
We also thank Tadashi Takayanagi for insightful comments on the draft.
We are grateful to the Isaac Newton Institute for Mathematical Sciences
for hospitality during the workshop on Mathematical Aspects of String and M-theory
where this work was started.
The work of S.C. has been supported by the Cambridge-Mitchell Collaboration in Theoretical
Cosmology, and the Mitchell Family Foundation.
The work of J.B. was supported by World Premier International Research Center Initiative (WPI Initiative), MEXT, Japan.

\appendix

\section{Restrictions from null-energy condition} \label{app:nullcond}

\FIGURE{
\centering
\includegraphics[width=\textwidth]{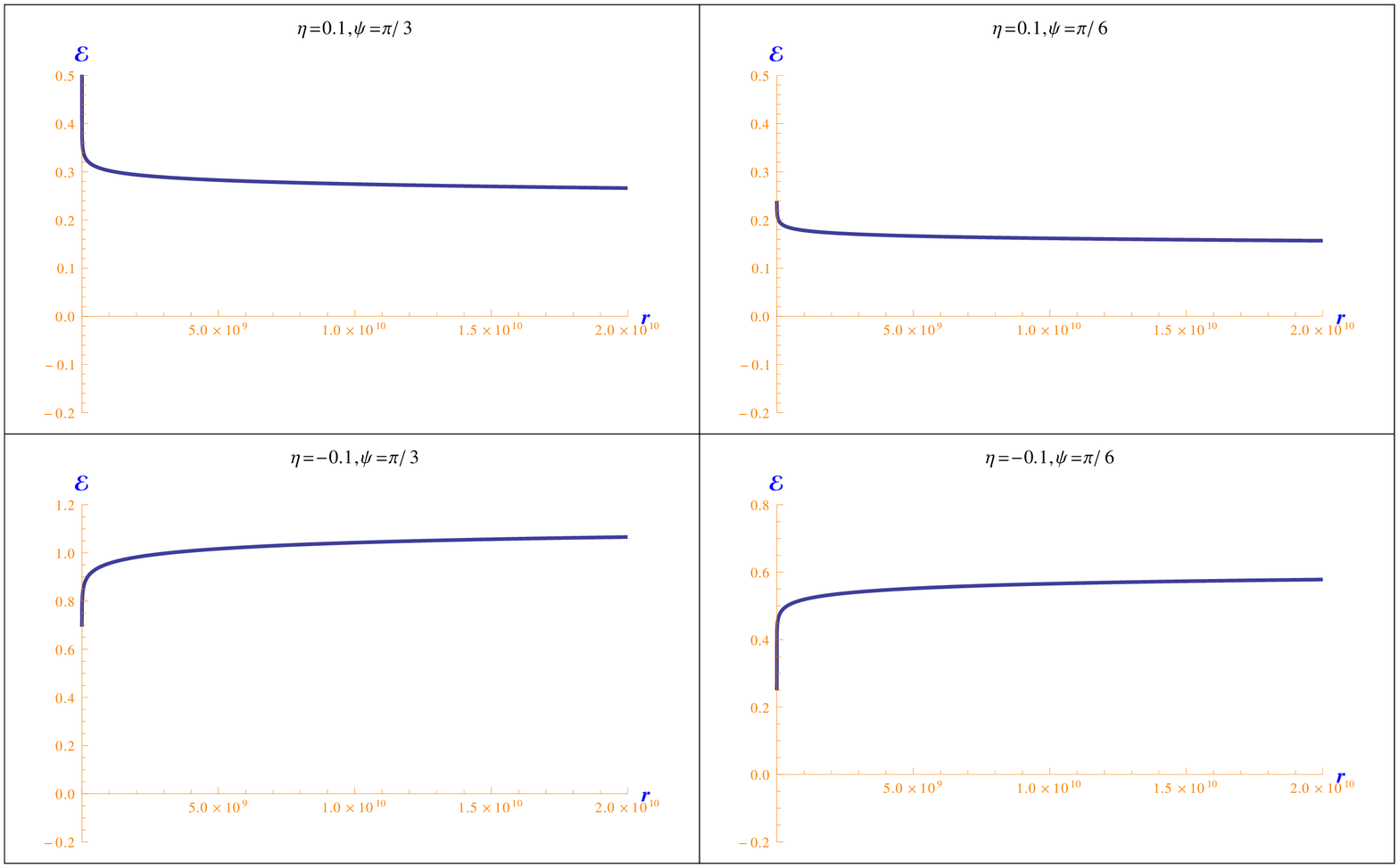}
\caption{Plots of ${\cal E}(r)$ for two distinct values of $\psi$ for the parameter set 1 ($\eta= 0.1$) and set 2 ($\eta$=-0.1). The fact that ${\cal E}(r)$ is positive for all
values of $r$ (for two distinct values of $\psi$) shows that the chosen set of parameters correspond to sensible matter. }
\label{fig:nulleng}
}

To test whether we have sensible matter in a theory with negative cosmological constant we generally impose the
null energy condition
\begin{equation}
 {\cal E} =n_{\mu} n_{\nu} T^{\mu \nu} \geq 0 \, ,
\end{equation}
where $n_{\mu}$ is an arbitrary null vector.
Here we would like to examine the restrictions which the null energy condition places on the structure of the hyperscaling violating
solutions to the classical theory given by (\ref{GeneralLag}) and (\ref{fVform}).
We choose the null vector to be of the form
\begin{equation}
 n^{\mu} = \{ \frac{1}{\sqrt{a(r)}}, \sqrt{a(r)} \cos \psi, \frac{1}{\sqrt{b(r)}} \sin \psi, 0 \} \, ,
\end{equation}
where $\psi$ is kept as an arbitrary parameter. Reading off the stress tensor from
the Eintein equation
\begin{equation}
 T_{\mu \nu} = R_{\mu \nu} -\half g_{\mu \nu} R \, ,
\end{equation}
we find  that for our metric ansatz \eqref{SolRepeat} $ {\cal E}$ evaluates to
\begin{equation}
 {\cal E} = \frac{1}{2} C_a^2 r^{-2 \gamma } \left(-4 \beta ^2+\left(\beta -2 \gamma ^2+3 \gamma -1\right)
 \cos (2 \psi )+3 \beta +2 \gamma ^2-3 \gamma +1\right) \, .
\end{equation}
For the null energy condition to hold for arbitrary $\psi$ we require
\begin{equation}
\label{NEC}
 \begin{split}
  \beta (\beta-1) \leq 0, \\
   (\beta+\gamma -1)(2 \beta -2\gamma +1) \leq 0.
 \end{split}
\end{equation}
The necessary and sufficient condition for the first condition to hold is $0 \leq \beta \leq 1$.
In addition to this, the second condition  constrains $\gamma$ to satisfy
$\gamma<1-\beta$ and $\gamma<\half + \beta$.
Given $0 \leq \beta \leq 1$ this implies that we must have $\gamma \leq 1$.
Note that the first null-energy condition in (\ref{NEC}) is equivalent to requiring that $K$ is real (recall that $K=\beta-\beta^2$).
For the case  $f(\phi)=e^{2\alpha\phi}$ which we have considered, the second null-energy condition turns out to be identical to
requiring that the magnetic charge (or more precisely, the quantity $C_a^2 \, C_b^4/Q_m^2$) is real, as can be seen from (\ref{Qm2C}).
Finally, note that $\gamma<0$ and $0 \leq \beta \leq 1$ satisfy automatically both null-energy conditions.

If we plot ${\cal E}$ for the numerical solutions that we obtained in \S \ref{num}, we arrive at what is shown
in fig. \ref{fig:nulleng}.
Although we don't expect the null energy condition to be satisfied in the system describing the deep infrared, in which we
are accounting for quantum corrections,
these plots (which show that ${\cal E}$ is everywhere positive) illustrate that the parameters we have chosen to glue
onto the (classical) hyperscaling violating solution correspond to sensible matter in the region where quantum corrections are negligible.



\newpage

\bibliographystyle{JHEP}
\bibliography{resolve_hyper}

\providecommand{\href}[2]{#2}\begingroup\raggedright\begin{thebibliography}{10}


\bibitem{Kachru:2008yh}
  S.~Kachru, X.~Liu and M.~Mulligan,
  ``Gravity Duals of Lifshitz-like Fixed Points,''
  Phys.\ Rev.\ D {\bf 78}, 106005 (2008)
  [arXiv:0808.1725 [hep-th]].

\bibitem{Taylor:2008tg}
  M.~Taylor,
  ``Non-relativistic holography,''
  arXiv:0812.0530 [hep-th].

\bibitem{Gubser:2009qt}
  S.~S.~Gubser and F.~D.~Rocha,
  ``Peculiar properties of a charged dilatonic black hole in $AdS_5$,''
  Phys.\ Rev.\ D {\bf 81}, 046001 (2010)
  [arXiv:0911.2898 [hep-th]].

\bibitem{Cadoni:2009xm}
  M.~Cadoni, G.~D'Appollonio and P.~Pani,
  ``Phase transitions between Reissner-Nordstrom and dilatonic black holes in 4D AdS spacetime,''
  JHEP {\bf 1003}, 100 (2010)
  [arXiv:0912.3520 [hep-th]].

\bibitem{Charmousis:2010zz}
  C.~Charmousis, B.~Gouteraux, B.~S.~Kim, E.~Kiritsis and R.~Meyer,
  ``Effective Holographic Theories for low-temperature condensed matter systems,''
  JHEP {\bf 1011}, 151 (2010)
  [arXiv:1005.4690 [hep-th]].
  
\bibitem{Perlmutter:2010qu}
 E.~Perlmutter,
``Domain Wall Holography for Finite Temperature Scaling Solutions,''
 JHEP {\bf 1102}, 013 (2011)
 [arXiv:1006.2124 [hep-th]].



\bibitem{Iizuka:2011hg}
  N.~Iizuka, N.~Kundu, P.~Narayan and S.~P.~Trivedi,
  ``Holographic Fermi and Non-Fermi Liquids with Transitions in Dilaton Gravity,''
  JHEP {\bf 1201}, 094 (2012)
  [arXiv:1105.1162 [hep-th]].




\bibitem{Gouteraux:2011ce}
  B.~Gouteraux and E.~Kiritsis,
  ``Generalized Holographic Quantum Criticality at Finite Density,''
  JHEP {\bf 1112}, 036 (2011)
  [arXiv:1107.2116 [hep-th]].
  
 \bibitem{Huijse:2011ef}
  L.~Huijse, S.~Sachdev and B.~Swingle,
  ``Hidden Fermi surfaces in compressible states of gauge-gravity duality,''
  Phys.\ Rev.\ B {\bf 85}, 035121 (2012)
  [arXiv:1112.0573 [cond-mat.str-el]].

\bibitem{Ogawa:2011bz}
  N.~Ogawa, T.~Takayanagi and T.~Ugajin,
  ``Holographic Fermi Surfaces and Entanglement Entropy,''
  JHEP {\bf 1201}, 125 (2012)
  [arXiv:1111.1023 [hep-th]].

\bibitem{Kulaxizi:2012gy}
  M.~Kulaxizi, A.~Parnachev and K.~Schalm,
  ``On Holographic Entanglement Entropy of Charged Matter,''
  arXiv:1208.2937 [hep-th].

\bibitem{Ammon:2012je}
  M.~Ammon, M.~Kaminski and A.~Karch,
  ``Hyperscaling-Violation on Probe D-Branes,''
  arXiv:1207.1726 [hep-th].
  
\bibitem{Cadoni:2012uf}
  M.~Cadoni and S.~Mignemi,
  ``Phase transition and hyperscaling violation for scalar Black Branes,''
  JHEP {\bf 1206}, 056 (2012)
  [arXiv:1205.0412 [hep-th]].

\bibitem{Perlmutter:2012he}
  E.~Perlmutter,
  ``Hyperscaling violation from supergravity,''
  JHEP {\bf 1206}, 165 (2012)
  [arXiv:1205.0242 [hep-th]].

\bibitem{Dey:2012rs}
  P.~Dey and S.~Roy,
  ``Intersecting D-branes and Lifshitz-like space-time,''
  Phys.\ Rev.\ D {\bf 86}, 066009 (2012)
  [arXiv:1204.4858 [hep-th]].

\bibitem{Dey:2012tg}
  P.~Dey and S.~Roy,
  ``Lifshitz-like space-time from intersecting branes in string/M theory,''
  JHEP {\bf 1206}, 129 (2012)
  [arXiv:1203.5381 [hep-th]].

\bibitem{Narayan:2012hk}
  K.~Narayan,
  ``On Lifshitz scaling and hyperscaling violation in string theory,''
  Phys.\ Rev.\ D {\bf 85}, 106006 (2012)
  [arXiv:1202.5935 [hep-th]].

\bibitem{Hartnoll:2012wm}
  S.~A.~Hartnoll and E.~Shaghoulian,
  ``Spectral weight in holographic scaling geometries,''
  JHEP {\bf 1207}, 078 (2012)
  [arXiv:1203.4236 [hep-th]].

\bibitem{Dong:2012se}
  X.~Dong, S.~Harrison, S.~Kachru, G.~Torroba and H.~Wang,
  ``Aspects of holography for theories with hyperscaling violation,''
  JHEP {\bf 1206}, 041 (2012)
  [arXiv:1201.1905 [hep-th]].

\bibitem{Li:2010dr}
  W.~Li and T.~Takayanagi,
  ``Holography and Entanglement in Flat Spacetime,''
  Phys.\ Rev.\ Lett.\  {\bf 106}, 141301 (2011)
  [arXiv:1010.3700 [hep-th]].
  
\bibitem{Li:2009pf}
  W.~Li, T.~Nishioka and T.~Takayanagi,
  ``Some No-go Theorems for String Duals of Non-relativistic Lifshitz-like Theories,''
  JHEP {\bf 0910}, 015 (2009)
  [arXiv:0908.0363 [hep-th]].


\bibitem{Horowitz:2011gh}
  G.~T.~Horowitz and B.~Way,
  ``Lifshitz Singularities,''
  Phys.\ Rev.\ D {\bf 85}, 046008 (2012)
  [arXiv:1111.1243 [hep-th]].

\bibitem{Bao:2012yt}
  N.~Bao, X.~Dong, S.~Harrison and E.~Silverstein,
  ``The Benefits of Stress: Resolution of the Lifshitz Singularity,''
  arXiv:1207.0171 [hep-th].

\bibitem{Goldstein:2009cv}
  K.~Goldstein, S.~Kachru, S.~Prakash and S.~P.~Trivedi,
  ``Holography of Charged Dilaton Black Holes,''
  JHEP {\bf 1008}, 078 (2010)
  [arXiv:0911.3586 [hep-th]].
  
\bibitem{Goldstein:2010aw}
  K.~Goldstein, N.~Iizuka, S.~Kachru, S.~Prakash, S.~P.~Trivedi and A.~Westphal,
  ``Holography of Dyonic Dilaton Black Branes,''
  JHEP {\bf 1010}, 027 (2010)
  [arXiv:1007.2490 [hep-th]].

\bibitem{Harrison:2012vy}
  S.~Harrison, S.~Kachru and H.~Wang,
  ``Resolving Lifshitz Horizons,''
  arXiv:1202.6635 [hep-th].

\bibitem{Kundu:2012jn}
  N.~Kundu, P.~Narayan, N.~Sircar and S.~P.~Trivedi,
  ``Entangled Dilaton Dyons,''
  arXiv:1208.2008 [hep-th].

\bibitem{Cadoni:2011nq}
  M.~Cadoni, S.~Mignemi and M.~Serra,
  ``Exact solutions with AdS asymptotics of Einstein and Einstein-Maxwell gravity minimally coupled to a scalar field,''
  Phys.\ Rev.\ D {\bf 84}, 084046 (2011)
  [arXiv:1107.5979 [gr-qc]].


\bibitem{Barisch:2011ui}
  S.~Barisch, G.~Lopes Cardoso, M.~Haack, S.~Nampuri and N.~A.~Obers,
  ``Nernst branes in gauged supergravity,''
  JHEP {\bf 1111}, 090 (2011)
  [arXiv:1108.0296 [hep-th]].

\bibitem{Donos:2011pn}
  A.~Donos, J.~P.~Gauntlett and C.~Pantelidou,
  ``Magnetic and Electric AdS Solutions in String- and M-Theory,''
  arXiv:1112.4195 [hep-th].
  

\bibitem{Donos:2011qt}
  A.~Donos, J.~P.~Gauntlett and C.~Pantelidou,
  ``Spatially modulated instabilities of magnetic black branes,''
  JHEP {\bf 1201}, 061 (2012)
  [arXiv:1109.0471 [hep-th]].
  


\end{thebibliography}\endgroup

\end{document}